\documentclass[prb,twocolumn,showpacs,preprintnumbers,superscriptaddress]{revtex4-1}
\usepackage[bookmarks=false, colorlinks=true, plainpages = true,linkcolor = blue, citecolor = blue, urlcolor = blue, filecolor =blue]{hyperref}

\usepackage{amsfonts}
\usepackage{graphicx}
\usepackage{amssymb}
\usepackage{latexsym}
\usepackage{psfrag}
\usepackage{dcolumn}
\usepackage{datetime}
\usepackage{multirow}
\usepackage{subfigure}
\usepackage{amssymb,rotate,color}
\usepackage{bm}

\begin{document}

\title[Low energy  geometries of copper nano-junctions exposed to water]
{Systematic study of low energy  geometries of copper nano-junctions exposed to water and to species that can result from dissociation of water}

%
\author{Firuz Demir}
\address{Science Department,  Spokane Community College, Spokane, WA, U.S.A.}
\author{Kevin Dean}
\address{Physics Department,  Khalifa  University  of Science  and  Technology,  P.O.  Box  127788, Abu Dhabi, UAE}

\author{George Kirczenow} 
\address{Department of Physics, Simon Fraser
University, Burnaby, B.C, Canada V5A 1S6}

\vspace{10pt}
 
\date{\today}

\begin{abstract}
A detailed computational analysis has been performed, 
considering  copper atomic  contacts that are exposed directly to water molecules, hydroxyl groups, and monatomic as well as molecular hydrogen and oxygen species.  
The optimized physical bonding structure, electrical conductance and inelastic tunneling spectra (IETS) have been determined theoretically for moderately large structures by performing appropriate ab-initio and semi-empirical calculations. 
By considering the aforementioned properties, it has been possible to determine that some of the molecular bridging structures may be regarded as being highly-probable outcomes, resulting from the exposure of copper electrodes to the atomic/molecular contaminants. 
We specifically identify the conductance properties of a variety of configurations including examples with very high and very low conductance values. This is done in order to identify junction geometries that may be realized experimentally and their conductance and IETS signatures. 
By reporting geometries with very high and very low conductance values here, we intend to provide a wider perspective view than previous studies of copper-molecular junctions that have focused on high conductance structures.  
In addition, we explore the properties of metal junctions with multiple molecules, a class of systems for which little theoretical work has been available in the molecular electronics literature. 
We find that water molecules surrounding the junction can influence the bonding geometry of the molecules within the junction and consequently can affect strongly the calculated conductances of such junctions. 

\end{abstract}
\maketitle

\section{Introduction}
While there is a large molecular electronics literature, most theoretical work on molecules in metal nanojunctions has focused on gold metal junctions containing a single molecule~\cite{Kirczenow-review2010}.  
Gold is an inert metal, and atoms or molecules therefore do not readily chemisorb onto a flat gold surface,~\cite{Nakazumi_JPCC_118_2014} notable exceptions being molecules with thiol~\cite{Kirczenow-review2010,LindsayRatner} and amine~\cite{Kirczenow-review2010,Venkataraman} end groups.
A property of high reactivity has also been reported in the case of Au nano particles on TiO$_2$, where oxidation of CO takes place~\cite{Haruta_ChemLett_1987}.
By contrast, copper atomic contacts exhibit an enhanced reactivity even toward hydrogen~\cite{Nakazumi_JPCC_118_2014}.

While Au molecular junctions have been studied extensively,~\cite{Kirczenow-review2010,LindsayRatner} copper molecular junctions have to date received much less attention. Thus, since copper has significantly
different chemical properties than gold, a systematic theoretical study of copper molecular junctions may be expected to be of fundamental interest.
The purpose of this article is to present the results of such a study. 
We consider copper atomic junctions that are exposed to water molecules,  hydroxyl groups, and monatomic as well as molecular hydrogen and oxygen species, and study moderately large copper junctions in the presence of both single and multiple atoms and molecules. While in most experiments on molecular junctions the junction has been exposed to many molecules, there  
has been little previous theoretical work on junctions with more than one molecule even for gold nanoelectrodes.~\cite{Emberly2001a,Emberly2001b}   
The results presented here show explicitly (see Fig.~\ref{FigLast}) that water molecules surrounding the junction can influence the bonding geometry of the molecules within the junction and consequently can affect strongly the calculated conductances of such junctions.

Fabrication of single-molecule junctions with a stable conductance is critical in the process of developing reliable single-molecule electronic devices. Unstable conductance value measurements are thought to be due to strong molecule-metal interactions~\cite{Kaneko_JCP_2010}. 
Various atomic junction configurations, with possible strong molecule-metal interactions,
have been previously studied computationally. 
Pt/H$_2$O single-molecular junctions have been studied by Tal {\em et al.}
\cite{Tal_Krieger_Leerink_Ruitenbeek_PRL_2008} 
 where  conductance enhancement and reduction were reported.  
By focusing specifically on the conductance values that are demonstrably above the experimentally detectable threshold region, reliable bench-marked experimental results have been reported~\cite{Li_Demir_PCCP_17_2015, Demir_Dean_JCP_2010}. 
In order to confidently identify the most probable junction geometries arising from strong molecule-metal interactions, studying small heteronuclear molecular junctions that are not well understood~\cite{Li_Demir_PCCP_17_2015}
(for example H$_2$O) is also critically important.

\section{Theoretical methods, results and discussion}
In the present work, we focus  on a variety of configurations including examples with very high and very low conductance values, in order to provide a wide perspective view for the reader. 
Without  promoting any specific geometry, we provide a rubric as a configuration selection guide for the experimentalists who are studying such structures  in order to assist with probable junction identification. 
The largest calculated conductance values correspond to junctions with direct bonding between metal atoms on opposite sides of the junction, a result consistent with physical intuition since metals are good conductors. 
The smallest conductance values are those for junctions where the electron's path between the metal electrodes passes through more than two water molecules in series. This does not conflict with the possible expectations since water is an insulator, its low conductance being due to the Fermi level of the electrodes being located in the HOMO-LUMO gap of the water molecules. 
Conductance values between these two extremes are determined by the complicated details of electron quantum tunneling via the atoms within and near the junction of the metal electrodes, and are sensitive to the specific details of the atomic geometry.

The H$_2$O dissociation reaction and proton relay reaction were studied with STM and first-principles calculations on Cu (110) surfaces~\cite{Kumagai_PRL_2008, Kumagai_Nat.Mater_2012}.
The dissociation reaction has been reported on Cu nano-particles
and Cu single crystal surface at high temperatures ($>$600 K)~\cite{Rodriguez_Chem.Int.Ed_2007, Nakamura_FaradayTrans_1990, Rodriguez_JPhysChemC_2009}, and it has been reported that the dissociation does not proceed at low temperature~\cite{Spitzer_Luth_SurfSci_1982}.~Experimental and theoretical studies of H$_2$O/Cu junctions have shown that formation of H$_2$O/Cu junctions at l0~K is highly probable, due to the stable conductance value of 0.1~g$_0$~\cite{Li_Demir_PCCP_17_2015}.  
Nevertheless, although the atomic configuration of hydrogen molecular junctions is not fully understood, the bridging of a single hydrogen molecule between Au electrodes has been clarified~\cite{Kiguchi_PhysRevB_2010}.
As a direct result of this, we realize the importance of considering whether or not the proposed conductance feature is necessarily due to the formation of a Cu atomic chain with adsorbed H$_2$O molecule(s). 
It is highly possible that the conductance feature may alternatively have been due to the disassociated (and subsequently mobile), fragments of H$_2$O molecule(s) in the junction. This possibility is one of the foci of our attention in this article. 
\begin{figure*}[ht!]
\centerline{\includegraphics[width=0.96\linewidth]{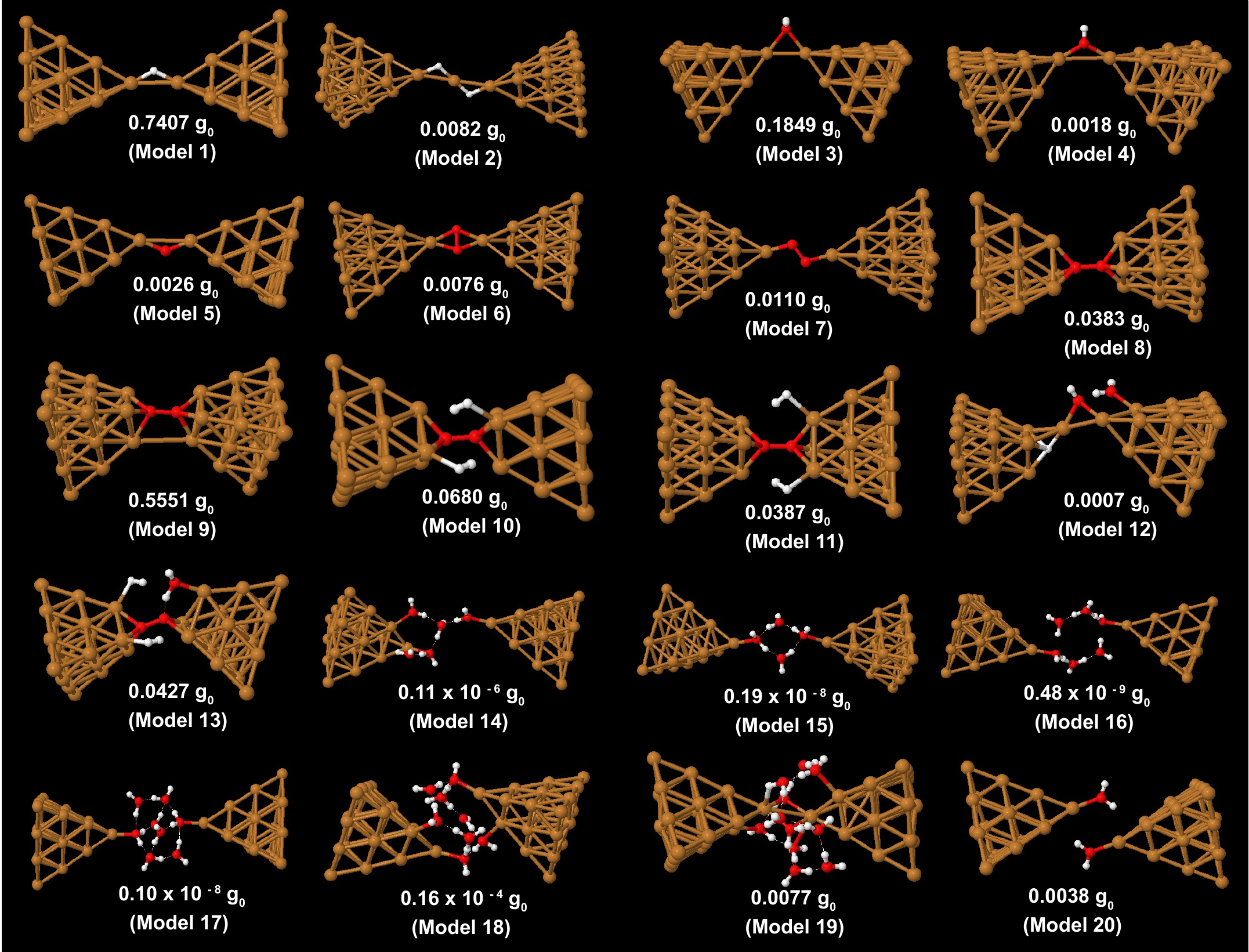}}
\caption{Energetically favorable states of possible junction types.  
Copper, hydrogen, and oxygen atoms are brown, white, and red, respectively.
These models are the computational  minimum energy configurations of the junctions which were selected as suitable candidates to optimize.}
\label{ModelsConsidered}
\end{figure*}

In order to simulate the stretching of the Cu junctions in relevant experiments, we 
initiated our calculations with a wide range of possible different trial bonding geometries, which were considered to be possible. 
We did not vary the junction size by stretching the Cu atomic clusters (that represent the electrodes) from the outer ends, but instead  simply held them fixed in the optimization process. Consequently, the individual components of the clusters forming the junction and molecule(s) bridging the junction were able to move freely in order to obtain the most computationally relaxed orientation.

 As an initial starting point, we imposed the condition that the water molecule(s) were in direct contact with the clusters. 
Occasionally we varied the bond lengths and/or bond angles of the water molecule(s) while they were still within the junction, in order to understand if they have a tendency to re-form or remain intact as molecules when computationally searching for the relaxed minimum energy configuration. 
Our {\em ab initio} calculations studying the junction formation indicated that there are minimum energy configurations for both possible scenaria, namely, for water molecule(s) that are intact in the junction and for water fragments that are in contact with the junction in various atomic or different molecular forms. We classified possible optimized junction geometries as depicted in Fig.\ref{ModelsConsidered}.

We carried out our optimizations at DFT level by using the PBE1PBE hybrid functional and Lanl2DZ basis sets with the GAUSSIAN'16 package~\cite{gaussian}. We have also repeated this for most cases with the ORCA package~\cite{Orca} using the same functional (PBE0) and the same Lanl2DZ basis sets. 
The results are very simillar. For example, the vibrational mode of Model 3 that found near 
264 cm$^{-1}$ (33 meV) with GAUSSIAN'16  is calculated to be at 263 cm$^{-1}$ (33 meV) with ORCA. 
The mode of Model 10 that is near 
339 cm$^{-1}$ (42 meV) according to GAUSSIAN'16  is at 
337 cm$^{-1}$ (42 meV) according to ORCA. 
The mode of Model 10 that GAUSSIAN'16 places near 
814 cm$^{-1}$ (101 meV)  is at 
801 cm$^{-1}$ (99 meV) according to ORCA. 
Because of the similarity we report here only the results obtained with GAUSSIAN'16.

Tunneling processes are sensitive to the interfaces, therefore, we utilized the same strategy as in our previous studies~\cite
{Li_Demir_PCCP_17_2015, 
Demir_Dean_JCP_2010, 
Li_Demir_PCCP_17_2015, 
Demir_Kirczenow_JCP_2011, 
Demir_Kirczenow_JCP_2012, 
Demir_Kirczenow_JCP_2012_2} 
to avoid computational potholes. Thus we calculated accurate equilibrium geometries of the complete molecule/electrode systems. 
In our optimization calculations with  the GAUSSIAN'16 package we included the D3 version of Grimme's empirical dispersion correction with Becke-Johnson damping (GD3BJ)~\cite{Grimme11}.
For calculations with the ORCA package, we also included 
the dispersion correction by using the key `D3BJ'. 
Once we had optimized our geometries, by using the optimized coordinates for the same functional and basis sets, we calculated the atomic displacements from equilibrium in the normal modes, as well as the corresponding frequencies with the packages mentioned above. 

The process of coupling these optimized 
structures to the macroscopic electron reservoirs and calculating the zero-bias tunneling conductances from the 
Landauer formula, 
\[  
g = g_0 \sum_{ij}    \vert
t_{ji}^{el}(\{{\bf 0}\}) \vert ^2 v_j/v_i~
\] 
where $g_0=2e^2/h$,  
are presented in 
Refs.~\cite{Demir_Kirczenow_JCP_2011, Demir_Kirczenow_JCP_2012, Demir_Kirczenow_JCP_2012_2}.  
These particular references also describe the solution of the Lippmann-Schwinger equation and evaluation of the appropriate Green's function by using semi-empirical extended H\"uckel theory \cite{yaehmop}, for the determination of the  elastic transmission amplitudes (conductance values).

By adopting the same theoretical approach as in Refs.~\cite{Demir_Kirczenow_JCP_2011, Demir_Kirczenow_JCP_2012, Demir_Kirczenow_JCP_2012_2}, 
we calculated IETS intensities,  
in order to hypothetically address which modes might have been giving rise to relatively higher, and therefore possibly experimentally detectable, conductance step heights upon applying the low bias voltages across the junction in the experiments.

Appropriate parametric values for the elastic transmission amplitudes are  shown in Fig.~\ref{ModelsConsidered}.  
The level of agreement between the theoretical and previously published experimental  values for `{\em probable}' geometries  was reported earlier in 
Refs.~\cite{Li_Demir_PCCP_17_2015, Demir_Dean_JCP_2010}, 
that were quoted as being: 
$0.2g_0$~\cite{Nakazumi_JPCC_118_2014}, 
$0.3g_0$~\cite{Yu_Li_JPCC_119_2015}
for Cu/H$_2$/Cu contacts, 
$0.1g_0$~\cite{Li_Demir_PCCP_17_2015, Yu_Li_PCCP2017} 
for Cu/H$_2$O/Cu contacts, 
and
$0.1g_0$~\cite{Yu_Li_JPCC_120_2016} 
for Cu/O$_2$/Cu contacts. 
Since we have meticulously followed the same methodology in our optimization and transport calculations, for the geometries in Fig.~\ref{ModelsConsidered},
obtaining low conductance measurements for these types of junctions is entirely plausible. 

When hydrogen and/or oxygen atoms form bonds in the junction (see Model 1-13 in Fig.~\ref{ModelsConsidered}), 
they act as impurities within the metallic crystal structure and consequently lower the conductance~\cite{Demir_Dean_JCP_2010}. 
The conductance changes due to bonding fragments of H$_2$O molecule(s) to the junction
may not be easily detectable in experiments; as seen in Fig.~\ref{ModelsConsidered} the conductance difference between 
Model 8 and Model 11 and/or Model 10 is not paramount,
consequently the signatures of their bonding geometries will not necessarily be readily identifiable.

\begin{figure}[h!]
\centerline{\includegraphics[width=1.0\linewidth]{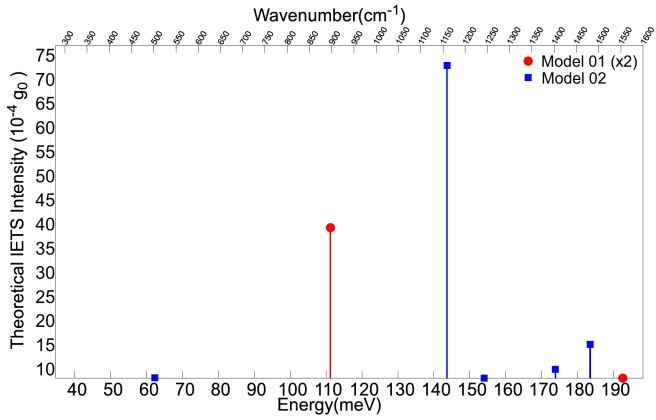}} 
\caption{Strong IETS intensities of Models of 1 and 2 are compared with respect to phonon energies.}
\label{Plot-01-02}
\end{figure}

\begin{figure}[h!]
\centerline{\includegraphics[width=1.0\linewidth]{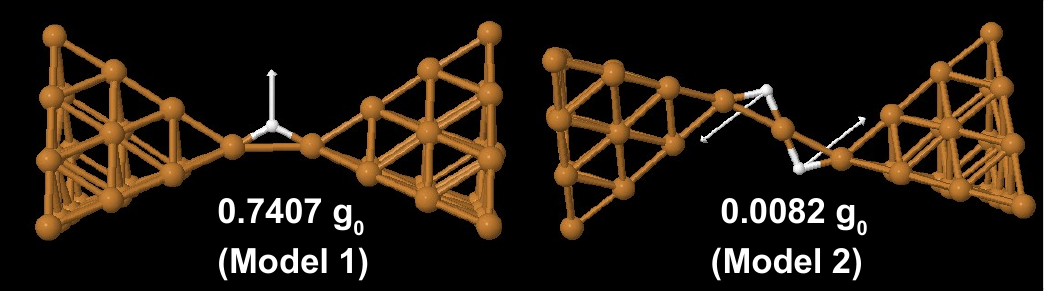}} 
\caption{Computational  minimum energy configurations of the junctions which were selected as suitable candidates to optimize.
Copper, hydrogen, and oxygen atoms are brown, white, and red, respectively. 
The arrows show the vibrational motion of the atoms for one of the strongest modes.}
\label{Plot-01-02+}
\end{figure}

In Fig.~\ref{Plot-01-02}, we have plotted the strong IETS intensity of Models 1 and 2 (of Fig.~\ref{ModelsConsidered}) with respect to their phonon energies. 
The video in the supplementary materials shows the atomic displacements of the vibrational modes that give rise to the features seen in the IETS spectra.
Note that Fig.~\ref{Plot-01-02+} shows atomic displacements with arrows for only one of the strongest modes, as an illustrative example.
%
%
%
%
%
Modes near 
111 meV 
(Model 1) and 
144 meV 
(Model 2) are very similar in motion where hydrogen is stretching back and forth perpendicular to the Cu-Cu bond in the interface. These are higher intensity modes. 
Modes near 
193 meV 
(Model 1) and 
184 meV 
(Model 2) are very similar in motion where hydrogen is stretching back and forth parallel to the Cu-Cu bond in the interface. 
However, the mode of Model 1 is very weak, and clearly not above the threshold level for suitable experimental detection. 

\begin{figure}[h!]
\centerline{\includegraphics[width=1.0\linewidth]{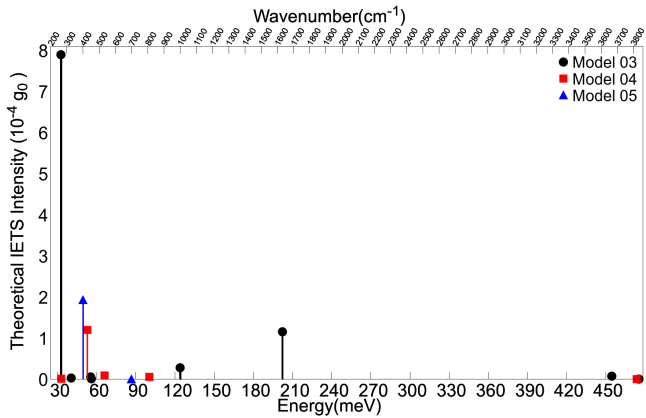}}
\caption{Strong IETS intensities of Models of 3, 4 and 5 are compared with respect to phonon energies.}
\label{Plot-03-04-05}
\end{figure}

\begin{figure}[h!]
\centerline{\includegraphics[width=1.0\linewidth]{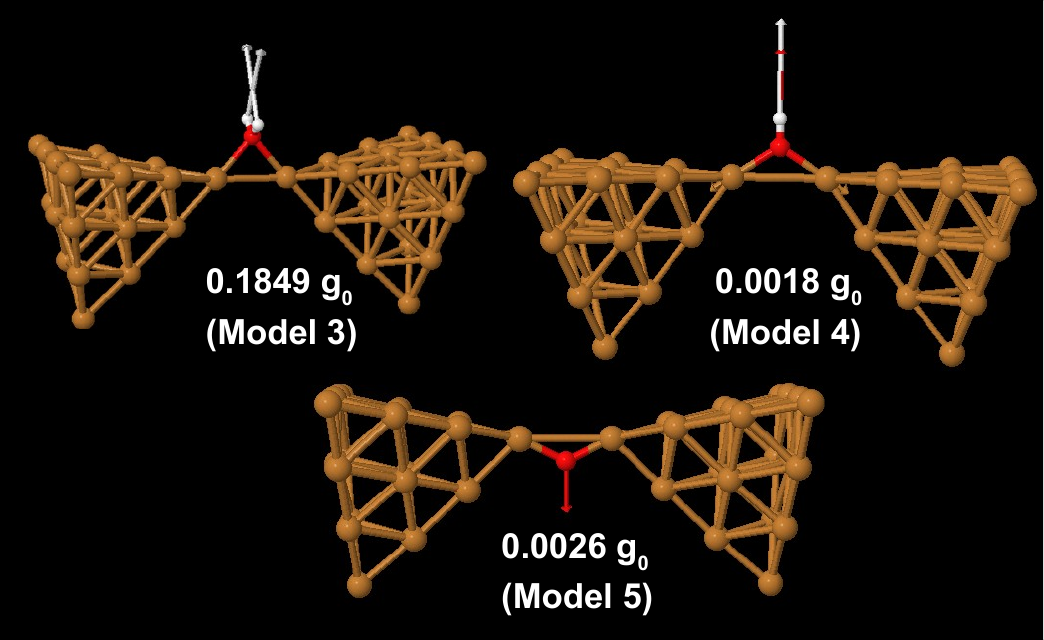}}
\caption{Computational  minimum energy configurations of the junctions which were selected as suitable candidates to optimize.
Copper, hydrogen, and oxygen atoms are brown, white, and red, respectively.
The arrows show the vibrational motion of the atoms for one of the strongest modes.}
\label{Plot-03-04-05+}
\end{figure}

In Fig.~\ref{Plot-03-04-05}, we have plotted the strong IETS intensity of Models 3, 4 and 5 (of Fig.~\ref{ModelsConsidered}) with respect to their phonon energies. 
The video in the supplementary materials shows the atomic displacements of the vibrational modes that give rise to the features seen in the IETS spectra.
Fig.~\ref{Plot-03-04-05+} shows this atomic displacements with arrows for only one of the strongest modes.
%
%
%
Modes near 
33 meV 
(Model 3), 
53 meV 
(Model 4) and
50 meV 
(Model 5) are breathing modes. 
Modes near 
55 meV 
(Model 3), 
66 meV 
(Model 4) and
87 meV 
(Model 5) are very weak modes. 
OH and O are stretching back and forth  in the junction parallel to the Cu-Cu bond in the interface. 
It is interesting to note here that a similar vibrational mode is absent in Model 3, where the water molecule is directly connected to the electrodes as a whole.

\begin{figure}[h!]
\centerline{\includegraphics[width=1.0\linewidth]{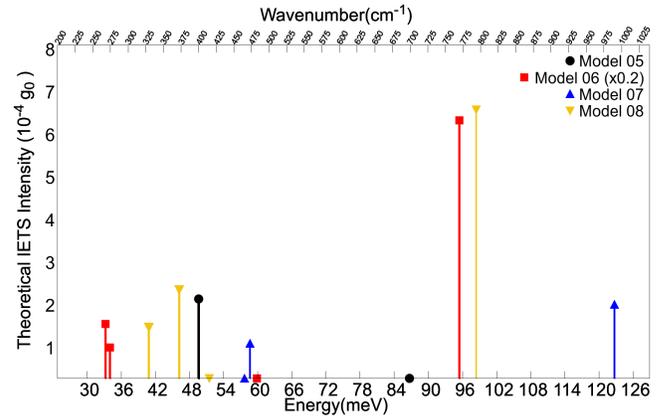}}
\caption{Strong IETS intensities of Models of 5, 6, 7 and 8 are compared with respect to phonon energies.}
\label{Plot-05-06-07-08}
\end{figure}

\begin{figure}[h!]
\centerline{\includegraphics[width=1.0\linewidth]{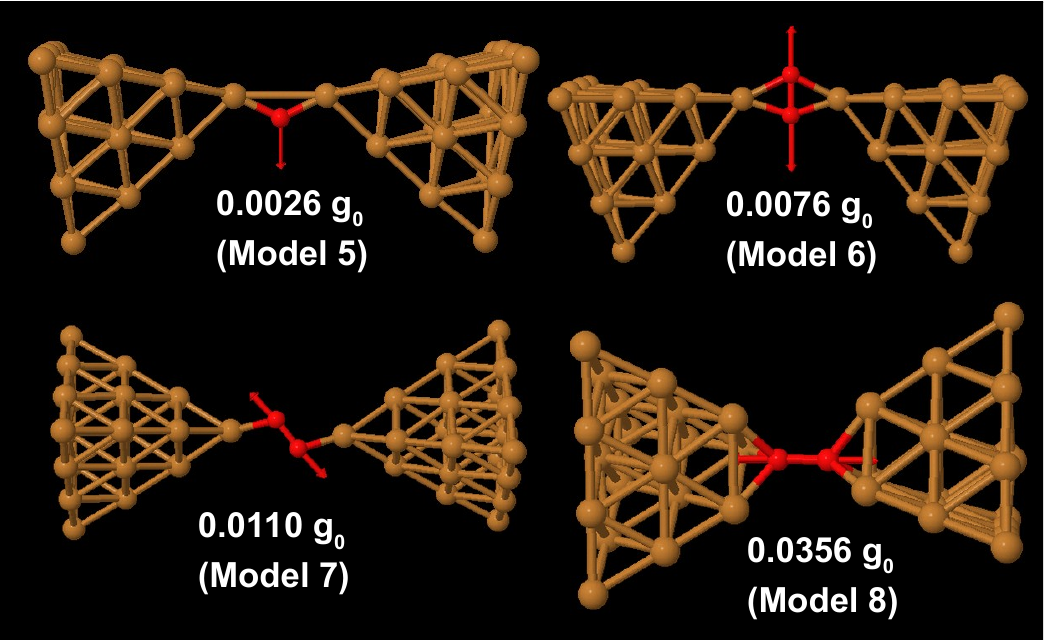}}
\caption{Computational  minimum energy configurations of the junctions which were selected as suitable candidates to optimize.
Copper, hydrogen, and oxygen atoms are brown, white, and red, respectively.
The arrows show the vibrational motion of the atoms for one of the strongest modes.}
\label{Plot-05-06-07-08+}
\end{figure}

In Fig.~\ref{Plot-05-06-07-08}, we have plotted the strong IETS intensity of Models 5, 6, 7 and 8 (of Fig.~\ref{ModelsConsidered}) with respect to their phonon energies.
The video in the supplementary materials shows the atomic displacements of the vibrational modes that give rise to the features seen in the IETS spectra.
Fig.~\ref{Plot-05-06-07-08+} shows atomic displacements with arrows for only one of the strongest modes.
%
%
Modes near 
50 meV 
(Model 5), 
34 meV 
(Model 6), 
59 meV 
(Model 7) and
46 meV 
(Model 8) are bending modes. 
Modes near 
87 meV 
(Model 5), 
95 meV 
(Model 6), 
123 meV 
(Model 7) and
98 meV 
(Model 8) are O-O stretching modes with higher intensities. 
We suggest that they should be easily detected in appropriate experiments.

\begin{figure}[h!]
\centerline{\includegraphics[width=1.0\linewidth]{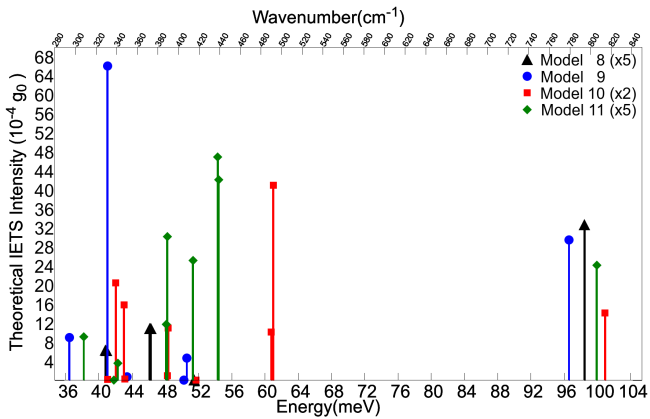}}
\caption{Strong IETS intensities of Models of 8, 9, 10 and 11 are compared with respect to phonon energies.}
\label{Plot-08-09-10-11}
\end{figure}

\begin{figure}[h!]
\centerline{\includegraphics[width=1.0\linewidth]{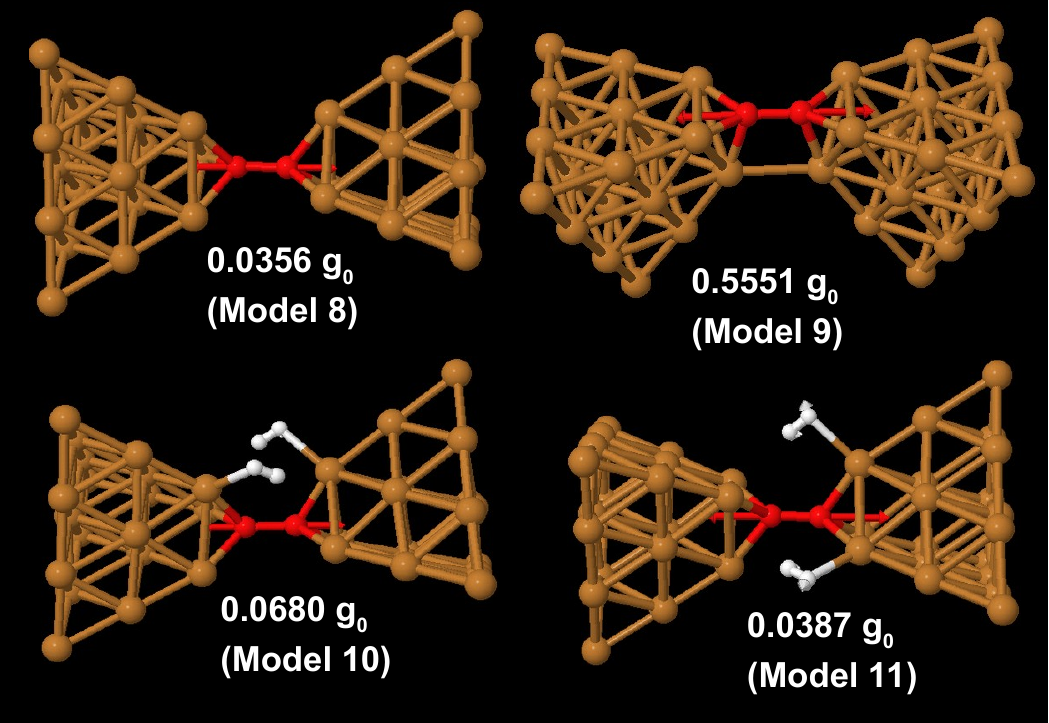}}
\caption{Computational  minimum energy configurations of the junctions which were selected as suitable candidates to optimize.
Copper, hydrogen, and oxygen atoms are brown, white, and red, respectively.
The arrows show the vibrational motion of the atoms for one of the strongest modes.}
\label{Plot-08-09-10-11+}
\end{figure}

In Fig.~\ref{Plot-08-09-10-11}, we have plotted the strong IETS intensity of Models 8, 9, 10 and 11 (of Fig.~\ref{ModelsConsidered}) with respect to their phonon energies.
The video in the supplementary materials shows the atomic displacements of the vibrational modes that give rise to the features seen in the IETS spectra.
Fig.~\ref{Plot-08-09-10-11+} shows atomic displacements with arrows for only one of the strongest modes.
%
%
%
We have magnified the vertical intensity scale of 
Mode 8 five times, 
Mode 10 two times and 
Mode 11 five times 
for clarity.
Modes near 
41 meV 
(Model 8),  
41 meV 
(Model 9) and 
42 meV 
(Model 10) 
 are strong perpendicular O-O wagging modes (refer to visualization), and should be readily detected in suitable experiments. 
Modes near 
61 meV 
(Model 10) and
54 meV 
(Model 11) 
are strong H-H bending modes. 
Modes near 
98 meV 
(Model 8),  
97 meV 
(Model 9), 
101 meV 
(Model 10) and
100 meV 
(Model 11) 
are O-O out-of-phase stretching modes. 
Based on their respective intensities, we suggest that they should be easily detected in appropriate experiments.
The animated visualization of these different types of motion can be found as a video in the supplementary materials.

\begin{figure}[h!]
\centerline{\includegraphics[width=1.0\linewidth]{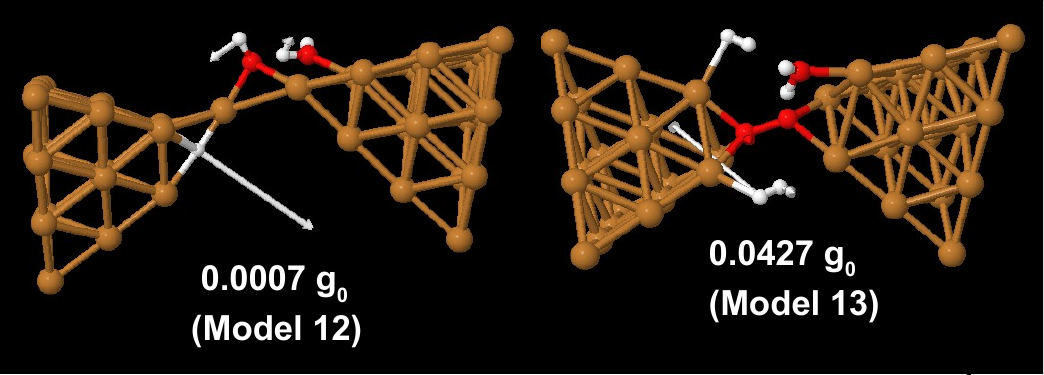}}
\caption{Computational  minimum energy configurations of the junctions which were selected as suitable candidates to optimize.
Copper, hydrogen, and oxygen atoms are brown, white, and red, respectively.
The arrows show the vibrational motion of the atoms for one of the strongest modes.}
\label{Plot-12-13+}
\end{figure}

In Fig.~\ref{Plot-12-13}, we have plotted the strong IETS intensity of Models 12 and 13 (of Fig.~\ref{ModelsConsidered}) with respect to their phonon energies. 
The video in the supplementary materials shows the atomic displacements of the vibrational modes that give rise to the features seen in the IETS spectra.
Fig.~\ref{Plot-12-13+} shows this atomic displacements with arrows for only one of the strongest modes.
%
\begin{figure}[h!]
\centerline{\includegraphics[width=1.0\linewidth]{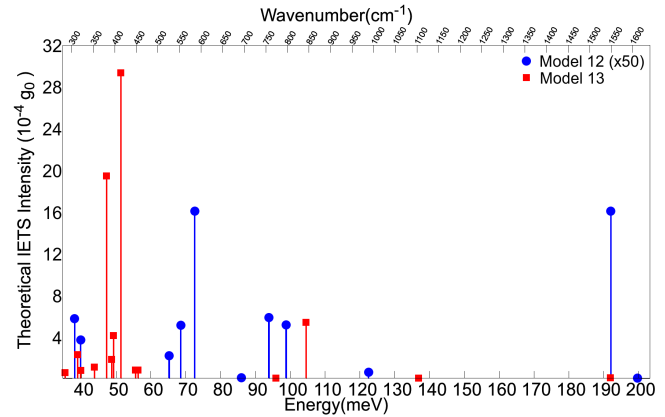}}
\caption{Strong IETS intensities of Models of 12 and 13 are compared with respect to phonon energies.}
\label{Plot-12-13}
\end{figure}
%
Those modes near 
73 meV, 
94 meV and  
99 meV and  
192 meV 
of Model 12 
are Cu-H stretching modes with higher intensities.
Modes near 
47 meV, 
51 meV and  
104 meV  
of Model 13 
are wagging, H-H bending and 0-0 stretching modes respectively. 
Although we are comparing these two modes here, 
it is not possible to match these modes exactly because their interface binding nature is  dissimilar.
This nature can be seen in the animated visualization video located in the supplementary materials.

In Fig.~\ref{Plot-14-15}, we have plotted the strong IETS intensity of Models 14, 15 (of Fig.~\ref{ModelsConsidered}), for comparison purposes. 
It is important to note that the vertical axis scale for Model 15 has been expanded by a factor of two.
\begin{figure}[ht!]
\centerline{\includegraphics[width=1.0\linewidth]{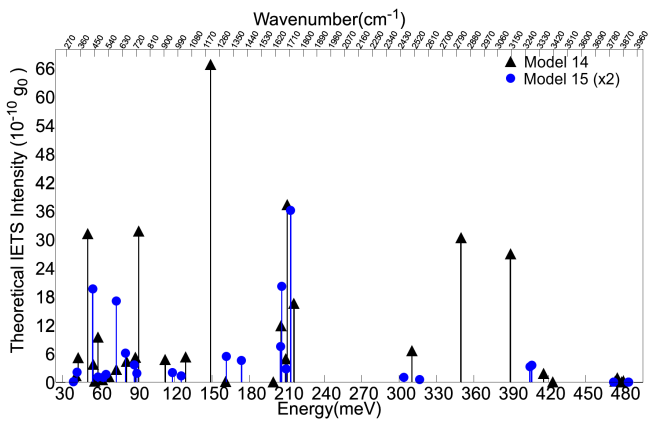}}
\caption{Strong IETS intensities of Models of 14 and 15 are compared with respect to phonon energies.}
\label{Plot-14-15}
\end{figure}

\begin{figure}[h!]
\centerline{\includegraphics[width=1.0\linewidth]{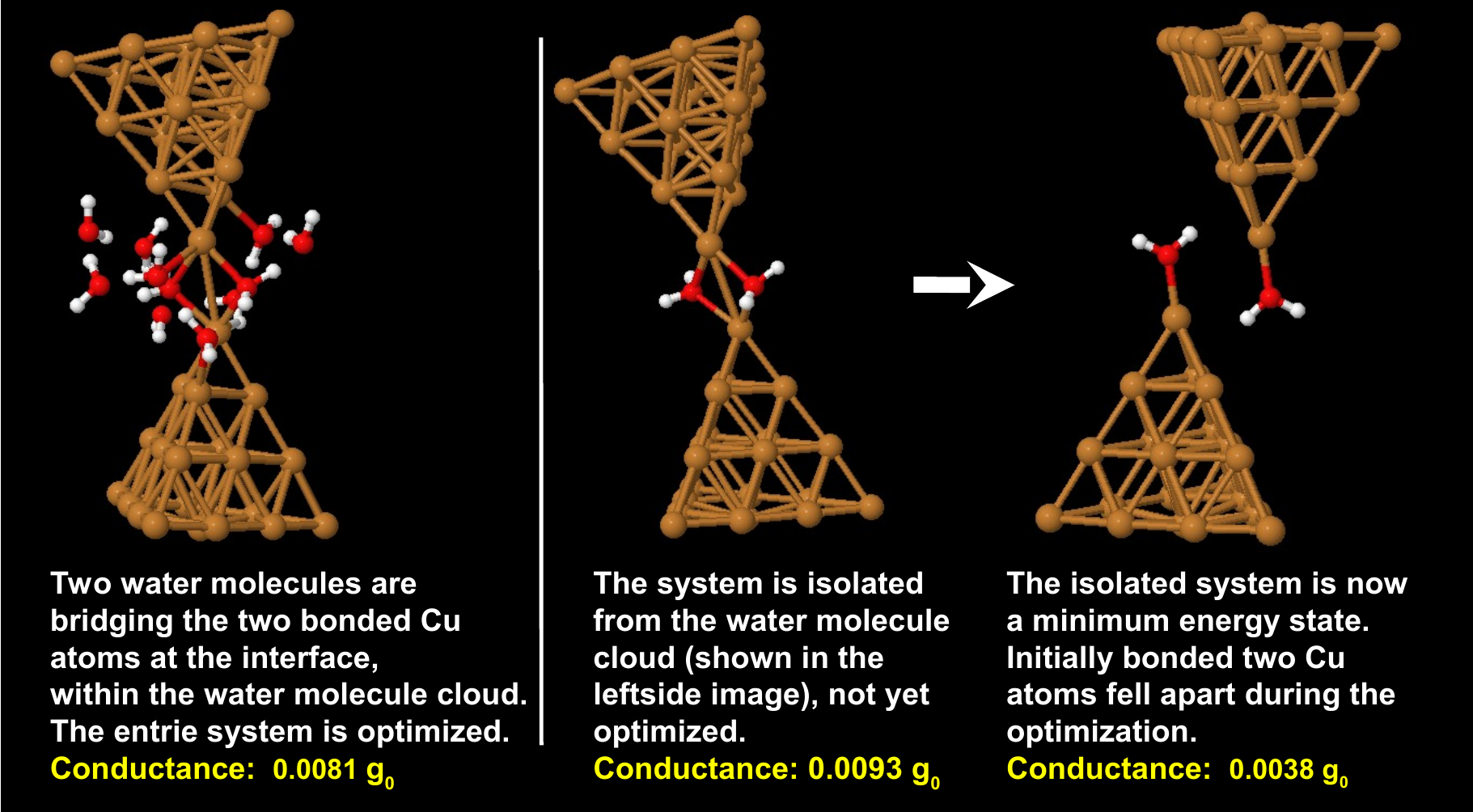}}
\caption{To be able to form junctions where two water molecules are bridging the junction side-by-side with the Cu-Cu direct bond/contact (on the left), we should not ignore the effect of the environment.}
\label{FigLast}
\end{figure}

In Fig.~\ref{FigLast}, we present a composite image of Models 19 and 20. 
At the center of the figure, 
 we show the copper junction with the two central molecules hypothetically isolated from the water molecule cloud, 
in a non-optimized molecular geometry, which is just one of the many possible intermediate transitional configurations between the core of Model 19 and Model 20. 
Once the structure at the center is completely optimized, it does not retain the two water and one Cu-Cu bridging nature, but instead it reforms into a geometry as shown at the right in Fig.~\ref{FigLast}. 
This clearly demonstrates that it is energetically unfavorable to form an isolated junction like the one at the center of Model 19 without the surrounding water molecule cage-like structure;  i.e., simulations that do not include the surrounding molecular cloud may yield geometries that are unlike those realized experimentally.

In Model 20 (shown in Fig.~\ref{ModelsConsidered}), it is important to note that although there is no direct molecular bond between the two oppositely aligned water molecules, 
it is hypothesized that quantum tunneling gives rise to the conductance. 
Although the conductance is still below the most recent experimentally detectable threshold region, it is noticeably not completely zero. 
%
%

\section{Conclusions} 
We have concentrated our attention in this article on the possibility that 
prominent conductance features reported in Cu/H$_2$O/Cu experiments may have been due to the disassociated (and therefore subsequently mobile), fragments of H$_2$O molecule(s) in the junction. 
Our calculations suggest that it is considered to be highly probable for water molecules, OH, monatomic as well as molecular hydrogen and also oxygen species, (that are exposed directly to atomic copper contacts) to contribute to the same conductance features that have been reported in experiments. 

It has been shown that a realistic problem cannot be adequately explained by appealing to typical over-simplified computer modeling.
Our results suggest that an isolated junction (meaning one without a surrounding water molecular cage) may not achieve the required energy minimization in order to form a conducting bridge, such as that which is achieved by the actual junction, which has the surrounding water molecules. 
We must therefore conclude that the physics of the electrical conductance process is subtly different in both cases. 
For example,  Model 19 and Model 20 both give rise to similar order of magnitude but have dissimilar conductance features. 

In summary,  we have performed a systematic computational analysis, which is intended to provide a wide perspective view for the reader, without promoting any specific geometry that might have been formed in most recent experiments and have provided a rubric as a configuration selection guide. 
However, in order to confidently identify the most probable junction geometry arising from molecule-metal interaction, we note that further detailed studies are still necessary.  

\vspace{2pc}
See {\em{Supplementary Material} }
for the animated visualization of different types of vibrational motion, which can be observed in the video.

\vspace{0pc}
The authors wish to acknowledge the invaluable contribution of the high-performance computing facilities at WestGrid and Compute Canada, and also the HPC cluster at the Masdar Institute in the UAE,  to the results of this research.
 This research was supported by NSERC.
We also thank A. Saffarzadeh  for helpful discussions. \\

\onecolumngrid
 \section*{References}
 \twocolumngrid


\begin{thebibliography}{99}

\bibitem{Kirczenow-review2010}For a review, see G. Kirczenow, \textit{Molecular nanowires and their properties as electrical conductors}, The Oxford Handbook of Nanoscience and Technology, Volume I: Basic Aspects, Chapter 4, edited by A. V. Narlikar and Y. Y. Fu, Oxford University Press, U.K. (2010).



\bibitem{Nakazumi_JPCC_118_2014} T. Nakazumi, S. Kaneko, and M. Kiguchi,     J. Phys. Chem. C {\bf 118}, 7489-7493 (2014).

\bibitem{LindsayRatner}S. M. Lindsay,  M. A. Ratner, Adv. Mater. {\bf 19}, 23-31 (2007).

\bibitem{Venkataraman}L. Venkataraman, J. E. Klare, I. W. Tam, C. Nuckolls, M. S. Hybertsen, M. L. Steigerwald,
Nano Lett. {\bf 6} 3458-462 (2006).


\bibitem{Haruta_ChemLett_1987}
M. Haruta, T. Kobayashi, H. Sano, N. Yamada, 
Chem. Lett., 405--408 (1987).

    
\bibitem{Emberly2001a}E. G. Emberly and G. Kirczenow, Phys. Rev. Lett. {\bf 87}, 269701 (2001).

\bibitem{Emberly2001b}E. G. Emberly and G. Kirczenow, Phys. Rev. B {\bf 64}, 235412 (2001).

\bibitem{Demir_Kirczenow_JCP_2012}
F. Demir and G. Kirczenow, J. Chem. Phys. {\bf 136}, 014703 (2012).
 
\bibitem{Demir_Kirczenow_JCP_2012_2}
F. Demir and G. Kirczenow, J. Chem. Phys. {\bf 137}, 094703 (2012).

\bibitem{Kaneko_JCP_2010}
S. Kaneko, T. Nakazumi and M. Kiguchi, J. Chem. Phys. {\bf 1}, 3520-3523 (2010)  

\bibitem{Tal_Krieger_Leerink_Ruitenbeek_PRL_2008} O. Tal, M. Krieger, B. Leerink and J. van Ruitenbeek, Phys. Rev. Lett, {\bf 100}, 196804 (2008).


\bibitem{Li_Demir_PCCP_17_2015}
Y. Li, F. Demir, S. Kaneko, S. Fujii, T. Nishino, A. Saffarzadeh, G. Kirczenow, and M. Kiguchi, 
Phys. Chem. Chem. Phys. {\bf 17}, 32436-32442 (2015). 



\bibitem{Demir_Dean_JCP_2010}
F. Demir and K. Dean, J. Chem. Phys. {\bf 150}, 024304 (2019) 


\bibitem{Kumagai_PRL_2008} T. Kumagai, M. Kaizu, S. Hatta, H. Okuyama, T. Aruga, I.
Hamada and Y. Morikawa, Phys. Rev. Lett., {\bf 100}, 166101 (2008).



\bibitem{Kumagai_Nat.Mater_2012}
T. Kumagai, A. Shiotari, H. Okuyama, S. Hatta, T. Aruga, I.
Hamada, T. Frederiksen and H. Ueba, Nat. Mater., {\bf 11}, 167-172 (2012)



\bibitem{Rodriguez_Chem.Int.Ed_2007}
J. A. Rodriguez, P. Liu, J. Hrbek, J. Evans, and M. Perez., Angew. Chem. Int. Ed., {\bf 46}, 1329-1332 (2007).





\bibitem{Nakamura_FaradayTrans_1990}
J. Nakamura, J. M. Campbell and C.T. Campbell, J. Chem. Soc., Faraday Trans., {\bf 86}, 2725-2734 (1990).





\bibitem{Rodriguez_JPhysChemC_2009}
J. A. Rodr\'iguez, J. Evans, J. Graciani, J. B. Park, P. Liu, J. Hrbek, and J. F. San, J. Phys. Chem. C., {\bf 113}, 7364-7370 (2009).



\bibitem{Spitzer_Luth_SurfSci_1982}
A. Spitzer and H. L\"{u}th., Surf. Sci., {\bf 120}, 376-388 (1982).



\bibitem{Kiguchi_PhysRevB_2010}
M. Kiguchi, T. Nakazumi, K. Hashimoto, K. Murakoshi, 
Phys. Rev. B, {\bf 81}, 045420/1−045420/4 (2010).


\bibitem{gaussian}
M. J. Frisch, G. W. Trucks, H. B. Schlegel, G. E. Scuseria, M. A.
Robb, J. R. Cheeseman, G. Scalmani, V. Barone, B. Mennucci, G. A.
Petersson \textit{et al.}, Gaussian 16 Revision: A.03, 
Gaussian, Inc., Wallingford, CT, 2016. 



\bibitem{Orca}
Frank Neese, Software update: the ORCA program system, version 4.0, Wiley Interdisciplinary Reviews: Computational Molecular Science, 8, 1, (2017).
 
\bibitem{Demir_Kirczenow_JCP_2011}
F. Demir and G. Kirczenow, J. Chem. Phys. {\bf 134}, 121103 (2011).


\bibitem{Grimme11}
S. Grimme, S. Ehrlich and L. Goerigk, 
Comp. Chem. {\bf 32}, 1456-65, (2011).




\bibitem{yaehmop}The version of extended H\"{u}ckel theory used 
was that of J. H. Ammeter, H.-B. B\"{u}rgi, J. C. Thibeault, and R. 
Hoffman, J. Am. Chem. Soc. {\bf 100}, 3686 (1978), as implemented in
the YAEHMOP numerical package by G. A. Landrum and W. V. Glassey
(Source-Forge, Fremont, California, 2001).



\bibitem{Yu_Li_JPCC_119_2015} Y. Li, S. Kaneko, S. Fujii, and M. Kiguchi,     J. Phys. Chem. C {\bf 119}, 19143-19148 (2015).


\bibitem{Yu_Li_PCCP2017} Y. Li, S. Kaneko, S. Fujii, T. Nishino, and M. Kiguchi, 
Phys. Chem. Chem. Phys. {\bf 19}, 4673 (2017). 




\bibitem{Yu_Li_JPCC_120_2016} Y. Li, S. Kaneko, Y. Komoto, S. Fujii, T. Nishino, and M. Kiguchi, J. Phys. Chem. C {\bf 120}, 16254-16258 (2016).

 




\end{thebibliography}
\end{document}